\documentclass[floatfix,prb,longbibliography,superscriptaddress,twocolumn,letterpaper,nobalancelastpage]{revtex4-2}

\usepackage{mycommands}

\begin{document}
	
	\title{Terahertz Landau level spectroscopy of Dirac fermions\\ in millimeter-scale twisted bilayer graphene}
	
	\author{Benjamin F. Mead}
	\affiliation{Department of Physics and Astronomy, University of Pennsylvania, Philadelphia, Pennsylvania 19104, USA}
	
	\author{Spenser Talkington}
	\email{spenser@upenn.edu}
	\affiliation{Department of Physics and Astronomy, University of Pennsylvania, Philadelphia, Pennsylvania 19104, USA}
	
	\author{An-Hsi Chen}
	\affiliation{Materials Science and Technology Division, Oak Ridge National Laboratory, Oak Ridge, Tennessee 37831, USA}
	
	\author{Debarghya Mallick}
	\affiliation{Materials Science and Technology Division, Oak Ridge National Laboratory, Oak Ridge, Tennessee 37831, USA}
	
	\author{Zhaodong Chu}
	\affiliation{Department of Physics and Astronomy, University of Pennsylvania, Philadelphia, Pennsylvania 19104, USA}
	\affiliation{Current Address: Materials Science Division, Argonne National Laboratory, Lemont, IL 60439, USA}
	
	\author{Xingyue Han}
	\affiliation{Department of Physics and Astronomy, University of Pennsylvania, Philadelphia, Pennsylvania 19104, USA}
	
	\author{Seong-Jun Yang}
	\affiliation{Department of Chemical Engineering, Pohang University of Science and Technology, Pohang 37673, Korea}
	\affiliation{Center for Van der Waals Quantum Solids, Institute for Basic Science, Pohang 37673, Korea}
	
	\author{Cheol-Joo Kim}
	\affiliation{Department of Chemical Engineering, Pohang University of Science and Technology, Pohang 37673, Korea}
	\affiliation{Center for Van der Waals Quantum Solids, Institute for Basic Science, Pohang 37673, Korea}
	
	\author{Matthew Brahlek}
	\affiliation{Materials Science and Technology Division, Oak Ridge National Laboratory, Oak Ridge, Tennessee 37831, USA}
	
	\author{Eugene J. Mele}
	\affiliation{Department of Physics and Astronomy, University of Pennsylvania, Philadelphia, Pennsylvania 19104, USA}
	
	\author{Liang Wu}
	\email{liangwu@sas.upenn.edu}
	\affiliation{Department of Physics and Astronomy, University of Pennsylvania, Philadelphia, Pennsylvania 19104, USA}
	
	\date{\today}
	
	\begin{abstract}
		Exotic electronic physics including correlated insulating states and fractional Chern insulators have been observed in twisted bilayer graphene in a magnetic field when the Fermi velocity vanishes, however a question remains as to the stability of these states which is controlled by the gap to the first excited state. Free-space terahertz magneto-optics can directly probe the gap to charge excitations which bounds the stability of electronic states, but this measurement has thus-far been inaccessible due to the micron size of twisted bilayer graphene samples, while the wavelength of terahertz light is up to a millimeter. Here we leverage advances in fabrication to create  twisted bilayer graphene samples over 5 mm $\times$ 5 mm in size with a uniform twist angle and study the magnetic field dependence of the cyclotron resonance by a complex Faraday rotation experiment in $p$-doped large angle twisted bilayer graphene. These measurements directly probe charge excitations in inter-Landau level transitions and determine the Fermi velocity as a function of twist angle.
	\end{abstract}
	
	\maketitle
	
	\section{Introduction}\label{section:intro}
	
	Over the past decade twisted-bilayer graphene (TBG) has cemented itself as the archetypal moir\'e material exhibiting superconductivity \cite{cao2018unconventional,yankowitz2019tuning}, correlated insulators \cite{cao2018correlated,lu2019superconductors}, fractional Chern insulators \cite{xie2021fractional} and a variety of other exotic electronic phases \cite{andrei2020graphene,zarenia2020enhanced} where the interplay of interactions, band geometry, and magnetic field dictate the electronic phase. In TBG, Fermi velocity renormalizes with twist angle \cite{dos2007graphene,neto2009electronic,shallcross2010electronic,rozhkov2016electronic} lead to the formation of flat bands at the magic angle of $\sim 1.1^\circ$ \cite{morell2010flat,trambly2010localization,bistritzer2011moire}. Recently it has been realized that correlated electron physics is not limited to the magic angle and that it can occur in Bernal bilayers where there have been observations of the quantum anomalous Hall effect \cite{geisenhof2021quantum}, superconductivity \cite{zhou2022isospin}, and correlated insulating states \cite{tsui2023direct}. This has spurred investigations looking for markers of correlation physics in the Landau level spectra by transport \cite{moriya2021probing} and infrared spectroscopy \cite{russell2023infrared} measurements of these bilayers. Key to the stability of all phases in bilayer graphene are the gap to excitations and the electronic dispersion (fixed by the value of the Fermi velocity in TBG): here we fabricate large size, large twist angle TBG samples and leverage free-space terahertz optics to directly probe the energy of inter-Landau level charge excitations and to determine the Fermi velocity.
	
	One of the most alluring aspects of TBG is its tunability by varying the twist angle, through doping, and by applying electric and magnetic fields. Within a region around the charge neutrality point bounded by van Hove singularities \cite{li2010observation,yan2012angle,wong2015local,kim2016charge}, TBG exhibits the linear dispersion of Dirac fermions for angles larger than the magic angle \cite{de2012numerical,shallcross2013emergent}, with a twist angle dependent Fermi velocity that completely specifies the low-energy electronic structure \cite{dos2007graphene}.
	Going beyond decoupled layers, interlayer coherence effects are important in large angle commensurate TBG structures such as the Bernal bilayer \cite{shallcross2008quantum,mele2010commensuration,mele2012interlayer} and the low-energy electronic structures are strongly tunable with an applied electric field \cite{mccann2006asymmetry,castro2007biased,mak2009observation,talkington2023electric,talkington2024linear}.
	Remarkably, even in the absence of time-reversal symmetry breaking from a magnetic field, TBG samples can exhibit a chiral optical response owing to natural optical activity from broken mirror-$z$ symmetry \cite{kim2016chiral,morell2017twisting,stauber2018chiral,addison2019twist,talkington2023terahertz} and in an applied magnetic field there can be giant Faraday rotation \cite{crassee2011giant}.
	
	Landau levels form in a magnetic field when electrons undergo cyclotron motion and form states with quantized energy and angular momentum in accordance with the Bohr-Sommerfeld quantization relation. For Dirac fermions, Landau levels are spaced as $\sqrt{B}$ as was demonstrated in scanning tunneling microscopy (STM) probes \cite{zhang2005experimental,novoselov2005two2,miller2009observing,song2010high} and infrared spectroscopy \cite{jiang2007infrared} of monolayer graphene shortly after it was first isolated \cite{novoselov2005two}. Large area epitaxial multilayer graphene, up to 100 layers thick, later reconfirmed the $\sqrt{B}$ dependence as stacking faults restored the monolayer-like electronic structure \cite{sadowski2006landau,orlita2008approaching,orlita2010dirac,witowski2010quasiclassical}. In tandem, Landau level spectroscopy of Bernal bilayers was deduced through careful transport measurements \cite{zou2011effective} and infrared spectroscopy of exfoliated bilayers \cite{henriksen2008cyclotron} and ``Bernal like" epitaxial multilayer graphene \cite{orlita2011magneto} in good agreement with theory \cite{mccann2006landau,abergel2007optical,cruise2024observability}.
	In contrast, twisted bilayer graphene has been inaccessible to free space terahertz probes because of the difficulty in producing samples with a uniform twist angle over a large length scale. Consequently other approaches have been taken to study the low energy electronic structure of TBG. To investigate the gap between Landau levels previous studies have used  scanning-tunneling microscopy \cite{song2010high,luican2011single,yin2015experimental,yin2015landau} and quantum oscillation measurements \cite{deheer2007epitaxial,rode2016berry,chung2018transport} on micrometer-size samples.
	
	Recent progress in ultra-flat substrates and wet-transfer techniques now allows for the preparation of TBG samples that satisfy the sample size requirements of free space terahertz probes \cite{yang2019all,yang2022wafer,yang2022twisted}. We leverage these techniques to fabricate millimeter-scale twisted bilayer graphene samples up to 5 mm $\times$ 5 mm in size with twist angles from $7^\circ$ to $20^\circ$ with up to a $1^\circ$ twist angle inhomogeneity. These samples are compatible with far-field terahertz spectroscopy, and we conduct time-domain magneto-terahertz polarimetry to resonantly excite inter-Landau level excitations and determine the Fermi velocity. The Fermi velocities we determine are consistent with previous STM and transport studies and thus provide a valuable benchmark of far-field terahertz spectroscopy on a known quantity. This develops terahertz spectroscopy into a tool to study the low energy electronic structure of two-dimensional moir\'e materials.
	
	\section{Cyclotron Resonances of Dirac Fermions in Twisted Bilayer Graphene}\label{section:cyclotron}
	
\begin{figure}
		\centering
		\includegraphics[width=\linewidth]{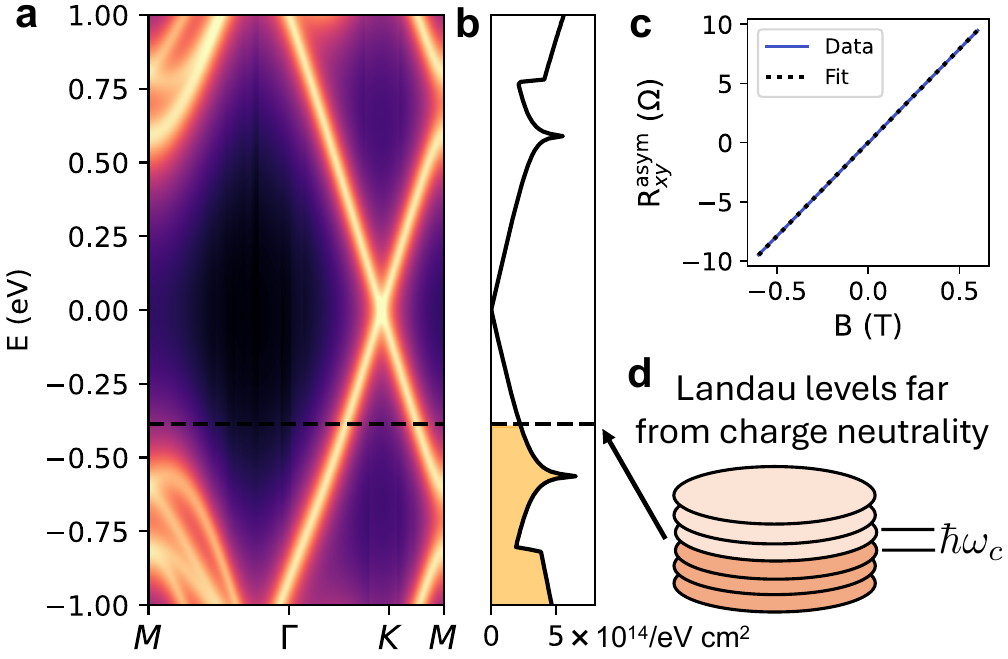}
		\caption{Electronic structure and doping of large angle twisted bilayer graphene. \textbf{(a)} Spectral function for a $9.43^\circ$ twist from a tight-binding model with phenomenological broadening $\eta=0.050$ eV, and \textbf{(b)} calculated density of states with chemical potential consistent with the $9^\circ$ experimental sample illustrated. \textbf{(c)} A linear fit to the asymmetric part of the measured Hall resistance in the $9^\circ$ sample gives the hole carrier density $n_h=3.966\times 10^{13}/\mathrm{cm}^2$. \textbf{(d)} Zooming in from the gross band structure to the fine structure of Landau levels the levels are separated by a cyclotron frequency $\hbar\omega_c$ in the terahertz range.}
		\label{fig:doping}
	\end{figure}

	The Landau energy levels for Dirac fermions are obtained by imposing the Bohr-Sommerfeld quantization rule, or equivalently by re-expressing the Hamiltonian in terms of raising and lowering operators \cite{luttinger1955motion}. Doing so one obtains \cite{sadowski2006landau}
	\begin{align}
		E_n = \text{sign}(n)\, \mathcal{E} \sqrt{|n|}
	\end{align}
	with $\mathcal{E}=v_F\sqrt{2e\hbar|B|}$. 
	Henceforth we shall assume $p$-doping so that $\sqrt{|n|}=\sqrt{-n}$. For low-lying Landau levels the cyclotron frequencies (gaps) are
	\begin{align}
		\hbar \omega_c^{-(n+1)\to -n} = \mathcal{E}(\sqrt{-n}-\sqrt{-(n+1)})
	\end{align}
	which exhibit a characteristic $\sqrt{B}$ dependence.
	Curiously for transitions between high lying Landau levels the cyclotron frequency is \textit{linear} in $B$ and forms a ``quasi-classical" limit as identified by Witowski et al in Ref. \cite{witowski2010quasiclassical}. This can be understood by noting that $\sqrt{n(n+1)}\approx n+\frac{1}{2}$ is a good approximation for sufficiently large $n$ so that
	\begin{align}\label{eq:cyclotron}
		\hbar \omega_c^{-(n+1) \to -n}
		&= \frac{\mathcal{E}}{\sqrt{|n|}}(\sqrt{n(n+1)}-n)
		\approx \frac{\mathcal{E}}{2\sqrt{|n|}}
		= \hbar \frac{eB}{\mu/v_F^2}
	\end{align}
	where we substituted $\mathcal{E}^2=2 e\hbar B v_F^2$ and $\sqrt{|n|}=\mu/\mathcal{E}$ where we can replace $E_n$ by $\mu$ since which Landau level is occupied is specified by the chemical potential. This cyclotron frequency $\hbar\omega_c$ is precisely in the terahertz range and can be resonantly excited, enabling terahertz spectroscopy of the Landau levels. We illustrate this schematically in Fig. \ref{fig:doping} where the chemical potential is far from charge neutrality yet Landau levels of Dirac fermions still form in this regime.
	
	Noting the similarity between Eq. \ref{eq:cyclotron} and the classical cyclotron frequency $\omega_c=eB/m^*$ we identify $\mu=m^*v_F^2$ which is precisely the Einstein rest-mass expression for the cyclotron mass in terms of the chemical potential and the Fermi velocity. Using this relation we can extract the Fermi velocity in terms of $m^*$ and $\mu$ which is expected to vary with twist angle as \cite{dos2007graphene}
	\begin{align}\label{eq:vF-expected}
		v_F(\theta) = v_F^\text{mono} \bigg[1 - 9 \bigg(\frac{t_\perp(\theta)}{\hbar v_F^\text{mono} \Delta K}\bigg)^2\bigg],
	\end{align}
	where $v_F^\text{mono}$ is the Fermi velocity of the monolayer, $t_\perp$ is the interlayer coupling, and $\Delta K$ is the magnitude of the separation of the wave-vectors of the Dirac points in the two layers.  
	
	\section{Experimental Study}\label{section:experiment}
	
	\begin{figure}
		\centering
		\includegraphics[width=\linewidth]{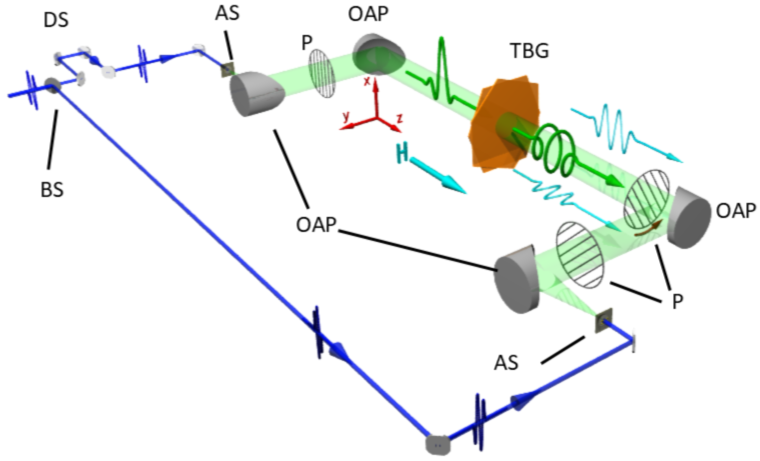}
		
		\caption{Illustration of the incident terahertz beam (green) on the TBG sample and the resulting Faraday rotation. Labeled optical elements include beam splitter (BS), Auston switch (AS), off-axis parabolic mirror (OAP) terahertz polarizer (P), and delay stage (DS).}
		\label{fig:faraday-setup}
	\end{figure}
	
	\begin{figure*}
		\centering
		\includegraphics[width=0.49\linewidth]{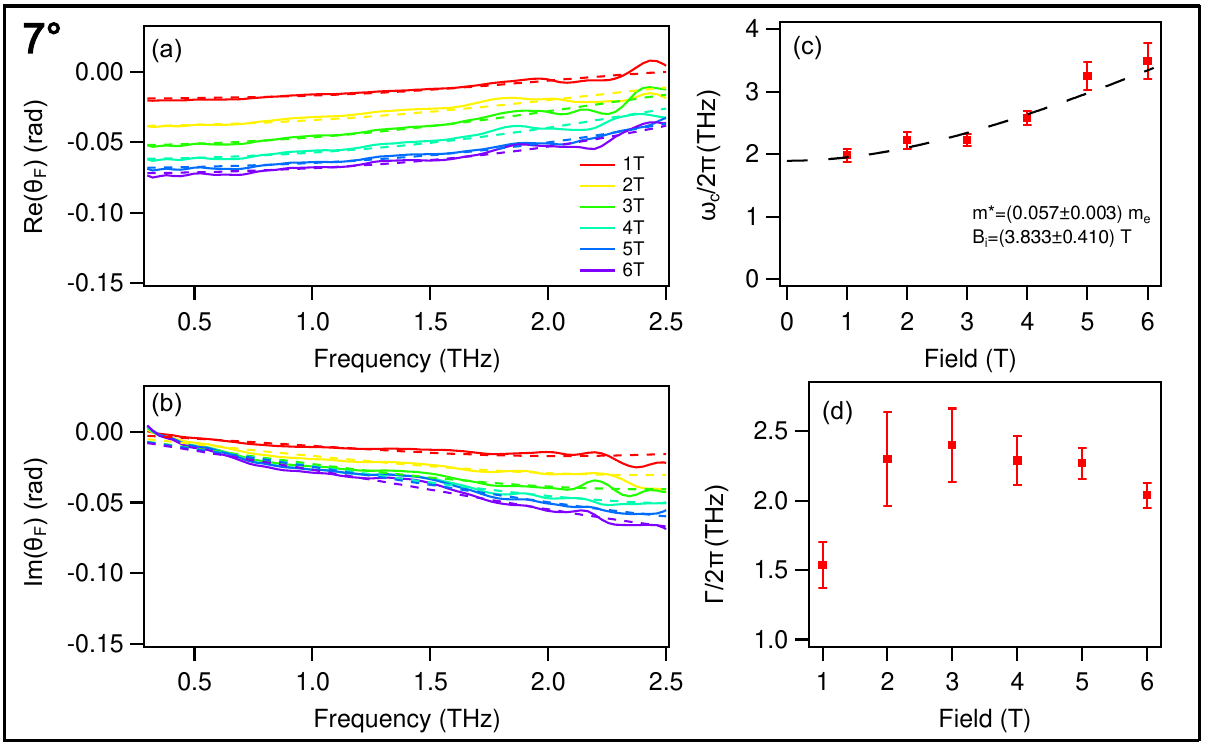}
		\includegraphics[width=0.49\linewidth]{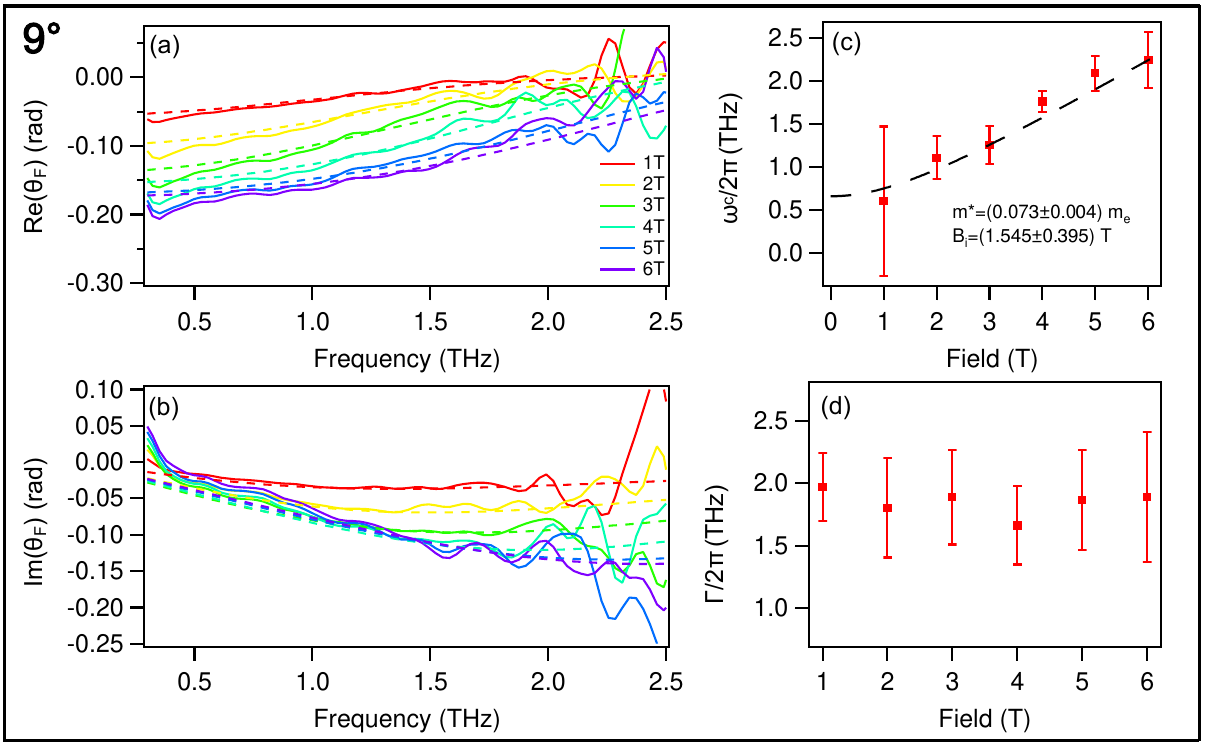}
		
		\includegraphics[width=0.49\linewidth]{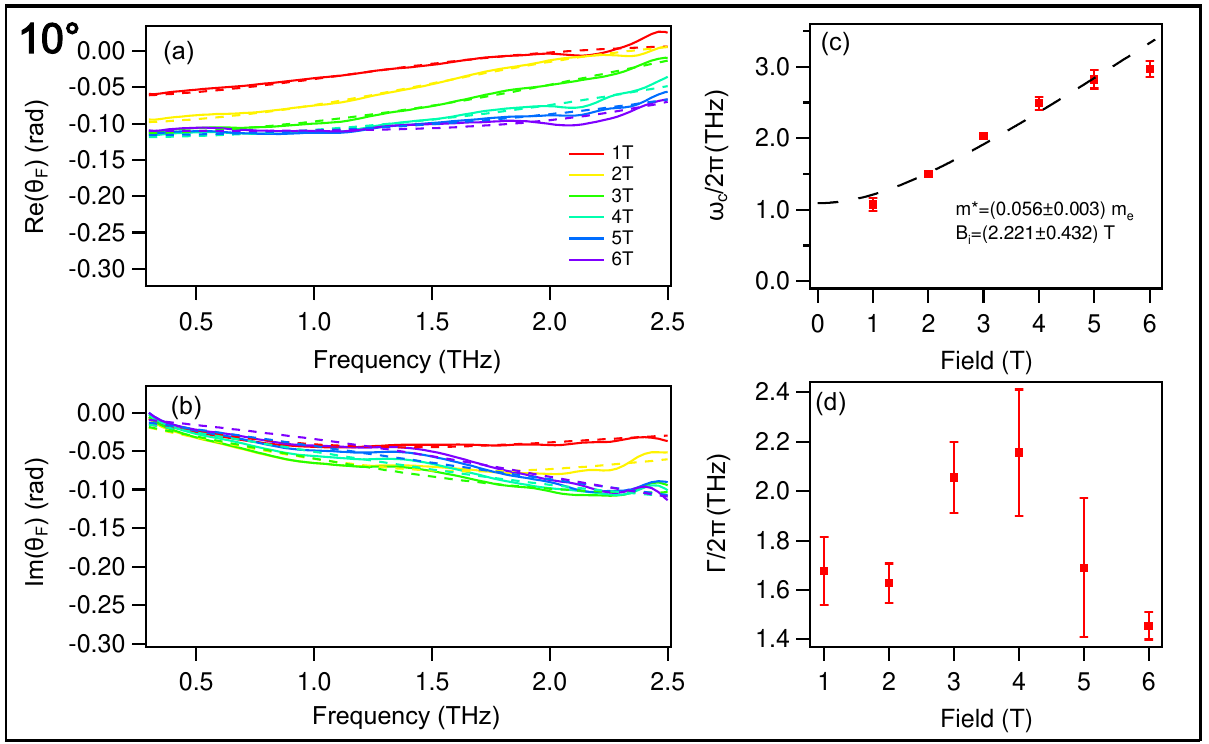}
		\includegraphics[width=0.49\linewidth]{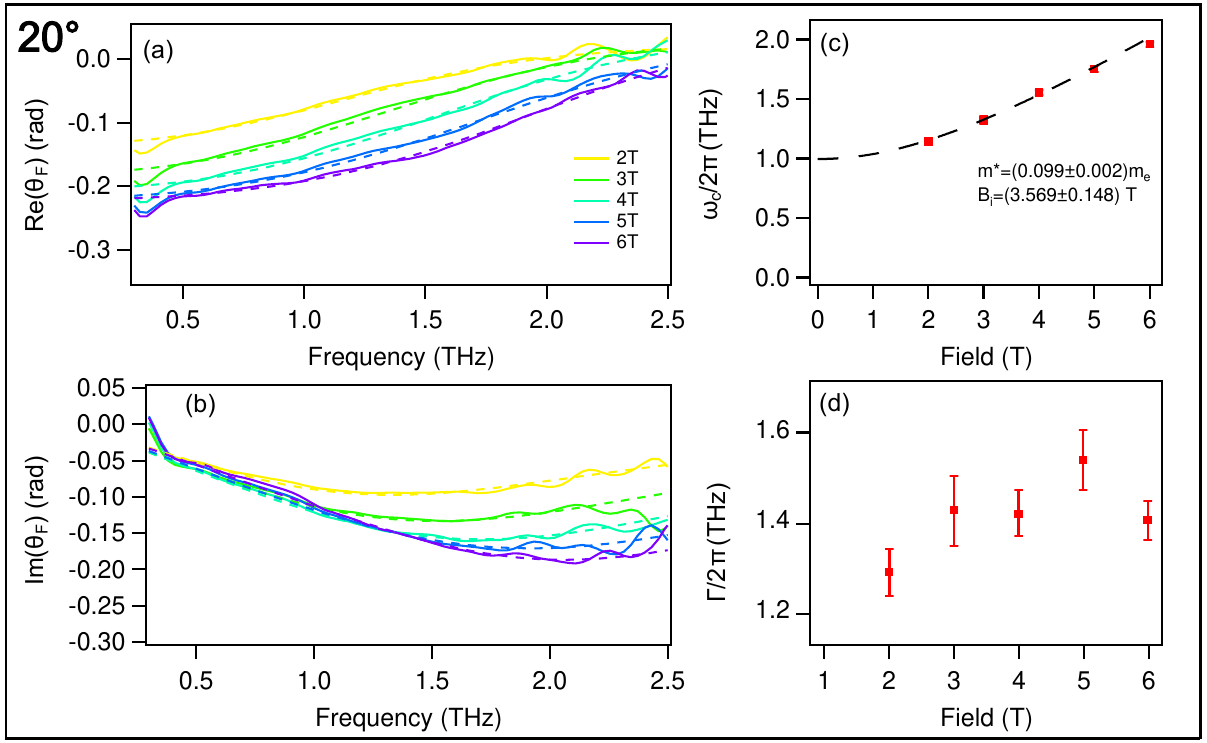}
		
		\caption{Magneto-terahertz measurements of $7^\circ$, $9^\circ$, $10^\circ$ and $20^\circ$ twisted bilayer graphene.  \textbf{(a)} Real and \textbf{(b)} imaginary parts of the Faraday rotation angle as dependent on magnetic field and light frequency; measured data is the solid lines and fits are the dashed lines. The samples are measured by terahertz pulses where the entire frequency range simultaneously probes the sample, however, the intensity of the pulse is much smaller at the extreme ends of the spectral range which results in noise at the upper end of the frequency range in the case $9^\circ$ sample. \textbf{(c)} Cyclotron frequency (gap between Landau levels) and \textbf{(d)} scattering rate extracted from fits to the experimental data; a fit to the formula $\omega_c=(e/m^*)\sqrt{B_i^2+B^2}$ gives the cyclotron mass $m^*$ and a length-scale $L_i=\sqrt{\hbar/eB_i}$ corresponding to disorder in the samples.}
		\label{fig:faraday}
	\end{figure*}
	
	To conduct Landau level spectroscopy, monolayer graphene was grown on germanium. The TBG samples were prepared from the monolayers with a previously developed wet transfer technique \cite{yang2022wafer} with the bilayers placed on a sapphire substrate.
	The Hall effect measurements were performed in a van der Pauw geometry with contacts made using pressed indium wires. The contacts were spaced roughly 1 mm apart to minimize any effects of tears which could be seen using an optical microscope. The measurements were performed in a Quantum Design Physical Property Measurement System at a base temperature of 2 K and field range up to 14 T.
	
	Next, the samples were placed on a 5 mm aperture for terahertz measurements. The time-domain terahertz setup is outlined in Fig. \ref{fig:faraday-setup}. A 780 nm wavelength femtosecond laser pulse (blue), 85 fs pulses, and 80 MHz repetition rate was split into to arms and sent to two photoconductive terahertz antennas. A delay stage changed the relative arrival time of the optical pulses on the terahertz antennas. The emitter antenna was biased to create current pulses that radiated into free space as broadband terahertz radiation. Four off-axis parabolic mirrors in an 8f focal geometry were used to direct the terahertz pulse to focus on the sample and then to the terahertz detector. When both the terahertz and optical pulse are incident on the detector antenna the produced current is linear in terahertz electric field and this was measured with a lock-in amplifier. Terahertz measurements were made in a Lakeshore cryostat at 1.7 K with a static magnetic field perpendicular to the surface of the sample.
	
	The frequency-dependent Faraday rotation angle was obtained by sending a terahertz pulse polarized in $\hat{x}$ to the sample and then separately measuring transmitted terahertz electric fields $E_{x}$ and $E_{y}$. Both electric fields are measured in the time domain and then Fourier transformed to obtain the spectrum. The spectrum was measured from 0.3 to 2.5 THz and was frequency limited by the spectrum of the initial terahertz pulse. The Faraday spectrum, $\theta_{F}(\omega)$, is then defined as 
	\begin{align}\label{eq:faraday_measure}
		\tan \left(\theta_{F}(\omega)\right)=\frac{E_{y}(\omega)}{E_{x}(\omega)}.
	\end{align}
	The Faraday spectrum is a complex quantity. The real component is a polarization rotation of the terahertz pulse, and the imaginary component describes the ellipticity of the transmitted pulse. The terahertz detector directly measures the terahertz electric field, not the intensity which is proportional to $E^{2}(\omega)$, so the real and imaginary Faraday angle was directly obtained without using Kramers-Kronig relations. 
	
	Our terahertz setup required three terahertz polarizers to measure $\theta_{F}$. One was placed before the sample to set the incident polarization. A second polarizer was placed on a motorized stage after the sample. This polarizer discriminates between the $\hat{x}$ and $\hat{y}$ components of the transmitted terahertz pulse. The third polarizer was set at $45^\circ$ right before the detector antenna. The terahertz antennas had a polarization preference, so the third polarizer made the detector equally sensitive to the $\hat{x}$ and $\hat{y}$ components \cite{han2022giant,han2024magneto}.    
	
	\section{Analysis}
	
	Using  the chemical potential from Hall effect measurements and the cyclotron mass extracted from terahertz Landau level spectroscopy we can determine the Fermi velocity. 
	
	\subsection{Chemical Potential}
	
	To obtain the chemical potential we obtained the carrier density from Hall effect measurements as described above. For our irregularly shaped samples $R_{xx}$ and $R_{xy}$ can be separated with van der Pauw symmetrization. Then the relation $n_h=R_{xy}^\mathrm{asym}(B)/eB$ where $R_{xy}^\mathrm{asym}(B)=\frac{1}{2}(R_{xy}(B)-R_{xy}(-B))$ gives the carrier density. We illustrate a representative measurement in Fig. \ref{fig:doping}(c). See Appendix \ref{section:hall} for full Hall effect data. Using the carrier density we then determined the chemical potential using theoretical models. We consider two models, first a Dirac-only model where the density of states is $D(E)=g_s g_v g_l E/\pi \hbar^2 v_F^2$ where $g$ are the spin, valley, and layer degeneracies and $v_F^2=m^*/\mu$. Integrating with $\mu$ an unknown quantity that can be obtained self-consistently
	\begin{align}
		n_h = \int_{\mu}^0 dE\ D(E) 
		= \frac{4 m^*}{\pi\hbar^2}\mu,
	\end{align}
	or $\mu = \pi\hbar^2 n_h/ 4m^*$ which is entirely in terms of measured quantities and natural constants. Second we use a tight-binding mode that accounts for the full lattice structure and long-range hopping. See Appendix \ref{section:TB} for details on the tight-binding model. We integrate the theoretical density of states until we reach the measured density. The density of states is the Brillouin zone integral of the single-particle spectral function
	\begin{align}
		A_{\bm{k},E} = - \frac{1}{\pi} \mathrm{Im}\bigg[\sum_n \frac{1}{E-\epsilon_n + i\eta}\bigg],
	\end{align}
	where $n$ runs over bands and $\eta$ is a phenomenological broadening.
	We illustrate the filling in Fig. \ref{fig:doping}(b) where the chemical potential is denoted with a dashed line. We see that the chemical potential is below the van Hove singularity, but far from charge neutrality Fig. \ref{fig:doping}(a).
	
	\subsection{Cyclotron Mass}
	
	The complex terahertz Faraday rotation, measured as illustrated in Fig. \ref{fig:faraday}, gives the cyclotron resonance which can be fitted to obtain the cyclotron mass. To analyze the Faraday rotation we follow the approach in Refs. \cite{hancock2011surface,valdes2012terahertz, wu2015high} where the complex Faraday rotation is defined as
	\begin{align}\label{eq:faraday}
		\theta_F = \tan^{-1}\left(-i \frac{t_+ - t_-}{t_+ + t_-}\right)
	\end{align}
	where $t_{\pm}$ ,the transmission for right/left-handed circularly polarized light, is
	\begin{align}
		t_\pm = \frac{2}{1+n_s} \frac{2n_s}{1 + n_s + Z_0 G_\pm}
	\end{align} 
	where $n_s$ is the substrate refractive index, $Z_0=4\pi\alpha \hbar/e^2\approx 376.73\ \Omega$ is the impedance of free space, and $G_\pm$ is the conductance under right/left-handed circularly polarized light. We take the conductance to be composed of a Drude-type term shifted by the cyclotron resonance and a lattice polarizability from absorptions above the measured spectral range
	\begin{align}\label{eq:G(w)}
		G_\pm(\omega,B) =i\epsilon_{0}\omega d\left(\frac{1}{\omega}\frac{\omega_{p}^{2}}{\omega\pm \omega_{c}(B)+i\Gamma} -(\varepsilon_{\infty}-1)\right)
	\end{align}
	where the term in parentheses is dimensionless, $d$ is the thickness of the bilayer, $\omega_p$ is the Drude plasma frequency, $\omega_{c}$ is the cyclotron resonance (gap between Landau levels), $\Gamma=2\pi/\tau$ is the scattering rate, $\tau$ is the average time between scattering events, and $\varepsilon_{\infty}$ is the lattice polarizability.  
	
	The measured Faraday rotations are the solid curves in Fig. \ref{fig:faraday}(a-b) and their fits obtained from Eq. \ref{eq:faraday} are the dashed lines. The cyclotron resonance appears as an inflection point in the real component of the Faraday spectrum and a minimum in the imaginary component. The 1 T data for the $20^\circ$ sample was omitted because the measured Faraday spectrum had oscillations that could not be fit with Eq. \ref{eq:G(w)}. Now, due to disorder effects, $\omega_c = eB/m^*$ is inaccurate at low fields. A fit to $\omega_c(B)$ that captures minimal features of disorder is the formula $\omega_c=\frac{e}{m^*}\sqrt{B_{i}^2+B^2}$ which has a non-zero asymptote corresponding to disorder scattering processes \cite{zaremba1991effect}. Fitting the cyclotron resonance as a function of magnetic field, Fig. \ref{fig:faraday}(c), gives the effective mass $m^*$ and impurity scattering length $L_i=\sqrt{\hbar/eB_i}$. We find $L_i$ of 13-21 nm for these samples, which is consistent with the length-scale for inter-valley scattering [74]. Fig. \ref{fig:faraday}(d) shows the scattering rate vs. magnetic field obtained from the fits.
	
	\subsection{Fermi Velocity}
	
\begin{figure}
		\centering
		\includegraphics[width=\linewidth]{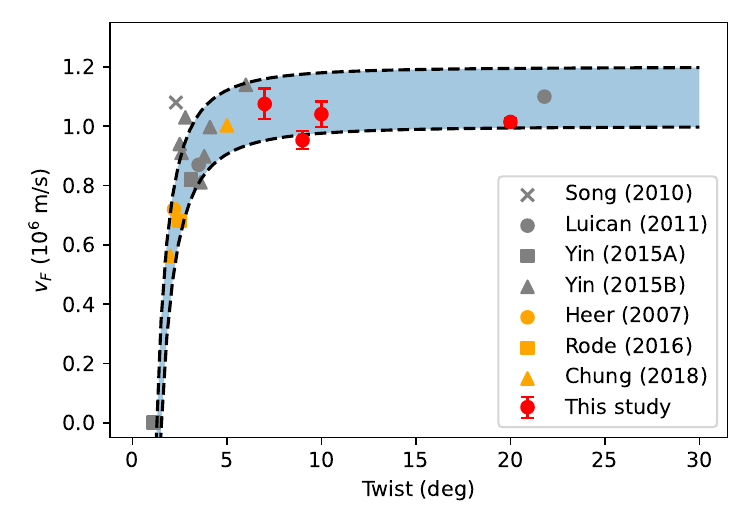}
		\caption{Fermi velocities determined by magneto-optical spectroscopy (red), previous STM studies (gray), and quantum oscillations studies (yellow). Lower and upper dashed lines are the theoretical fit from Eq. \ref{eq:vF-expected} with $v_F^\mathrm{mono}=1.0$ and $1.2\times 10^6$ m/s respectively and $t_\perp = 0.1$ eV. Variation in monolayer Fermi velocity can originate from the level of doping \cite{elias2011dirac,hwang2012fermi}.}
		\label{fig:velocities}
	\end{figure}

	We obtain the Fermi velocity by combining the transport and terahertz results. For TBG we plot the velocities in Fig. \ref{fig:velocities} and see that all the samples are consistent with the relatively constant Fermi velocity for $\theta\gtrsim 5^\circ$ predicted by theory and are in the range of Fermi velocities previously determined using STM and transport probes of small-scale samples. Perhaps somewhat surprisingly the Fermi velocities we determine for all samples are in the expected range despite significant $p$-doping; in particular the chemical potential of the $7^\circ$ sample lies near the van Hove singularity so we might expect a breakdown of Dirac physics for this sample. Now if twist angle was the only difference between samples we would expect the $9^\circ$ and $10^\circ$ to be nearly indistinguishable, however our samples have different chemical potentials, mobilities, and disorder profiles leading to the observed Fermi velocities. For extended data see Appendix \ref{section:data}. Our results demonstrate a new experimental approach to probe the electronic structure of two-dimensional materials and we find that Dirac-like physics persists well away from charge neutrality in twisted bilayer graphene.
	
	\section{Discussion}\label{section:conclusion}
	
	Two dimensional materials have emerged as compelling platforms to realize correlated electronic physics. Until recently this physics has been limited to short length scales of $\sim 10\ \mu\mathrm{m}$, but with advances in materials synthesis, twisted bilayer graphene with a uniform twist angle on the length scale of $\sim 1000\ \mu\mathrm{m}=1 \ \mathrm{mm}$ is now available. Leveraging these techniques we interfaced millimeter length scale twisted bilayer graphene with terahertz magneto-optics and conducted complex Faraday rotation and transport measurements to reveal the electronic structure of these materials. We measure the gap between Landau levels and find Fermi velocities that are consistent with previous scanning tunneling microscopy and transport studies thus validating our far-field approach on large samples. This suggests that these macroscopic samples exhibit good coherence over long length scales, a necessary prerequisite to realize reproducible macroscopic quantum materials. Extending bilayer materials from microscopic to mesoscopic and macroscopic length scales enhances scalability for advancing quantum technologies and may enable the realization of other exotic phenomena such as weak localization and universal conductance fluctuations \footnote{S. Talkington, D. Mallick, A.H. Chen, B. F. Mead, S. J. Yang, C. J. Kim, S. Adam, L. Wu, M. Brahlek, and E. J. Mele, \textit{Weak localization in twisted bilayer graphene} (in prep.)}.
	Looking ahead, terahertz techniques are a promising direction to probe collective excitations in two-dimensional materials such as twisted bilayer graphene and transition metal dichalcogenides.\\
	
	\section{Acknowledgments}
	The terahertz measurements were mainly supported by L.W.'s startup package at Penn. B.M. was partially sponsored by the Army Research Office under Grant Number W911NF-20-2-0166 and W911NF-25-2-0016. S.T. acknowledges support from the NSF under Grant No. DGE-1845298. A.H.C., D.M., and M.B. were supported by the U. S. Department of Energy (DOE), Office of Science, Basic Energy Sciences (BES), Materials Sciences and Engineering Division, and the National Quantum Information Science Research Centers, Quantum Science Center. X.H. acknowledges the partial support by Air Force Office of Scientific Research under award no. FA955022-1-0410 and the NSF EPM program under grant no. DMR-2213891. S.J.Y. and C.J.K. acknowledge support from the Institute for Basic Science (IBS), Korea, under Project Code IBS-R034-D1. E.J.M. was funded by the Department of Energy under grant DE-FG02-84ER45118. L.W. acknowledges support from the Sloan Foundation under the award FG-2025-25036.
	
%

	\clearpage
	\appendix
	
	\section{Hall Effect Analysis}\label{section:hall}
	
	Here we plot the Hall effect transport data and the linear fits $R_{xy}^\mathrm{asym}=B/n_he$ in Fig. S1.
	
	\begin{center}
		\centering
		\includegraphics[width=0.48\linewidth]{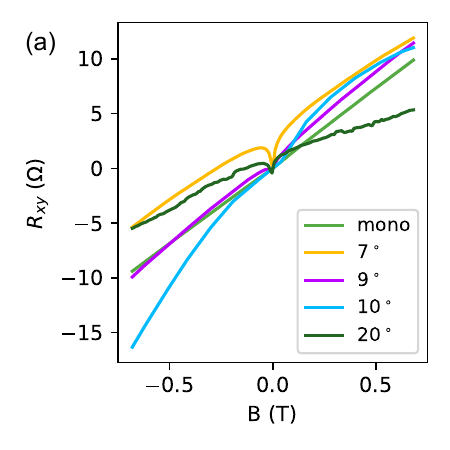}
		\includegraphics[width=0.48\linewidth]{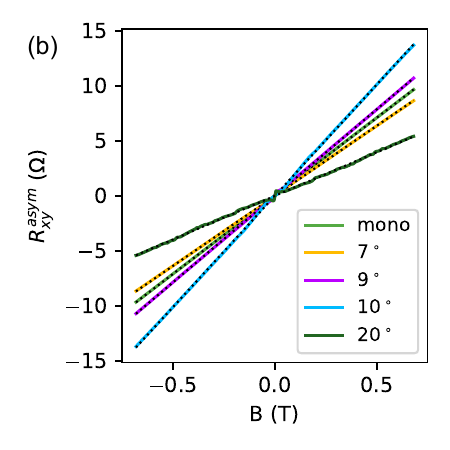}
		
		\small
		FIG. S1. Hall resistance from transport measurements at $T=2$ K. \textbf{(a)} Raw resistance data; shifted by a constant offset so that $R_{xy}(0)=0$. \textbf{(b)} Asymmetric part of $R_{xy}$ given by $R_{xy}^\mathrm{asym}(B)=\frac{1}{2}(R_{xy}(B)-R_{xy}(-B))$, and linear fit to $R_{xy}^\mathrm{asym}=B/n_h e$ where 1T/1$\Omega\cdot e=6.242 \times 10^{14}/\mathrm{cm}^2$.
		\label{fig:hall}
	\end{center}
	
	\section{Tight-Binding Results}\label{section:TB}
	
	We use a Slater-Koster tight-binding model of $p$ and $s$ orbitals on carbon atoms with hopping $t(\bm{r}) = V_{pp\pi}(\bm{r}) + V_{pp\sigma}(\bm{r})$ developed and used in Refs. [\onlinecite{trambly2010localization}] and [\onlinecite{moon2013optical}], where the hopping terms are given by
	\begin{align}
		V_{pp\pi}(\bm{r}) &= t_{pp\pi} e^{-(|\bm{r}|-a_0)/\delta} \left(1-\left(\frac{\bm{r}\cdot \bm{e}_z}{|\bm{r}|}\right)^2\right)\\
		V_{pp\sigma}(\bm{r}) &= t_{pp\sigma} e^{-(|\bm{r}|-d)/\delta}\left(\frac{\bm{r}\cdot \bm{e}_z}{|\bm{r}|}\right)^2
	\end{align}
	for hopping amplitudes $t_{pp\pi} = -2.7$ eV and $t_{pp\sigma} = 0.48$ eV, unit vector perpendicular to the bilayer $\bm{e}_z = (0,0,1)$, intra-layer atomic spacing $a_0 = a/\sqrt{3} = 0.142$ nm, inter-layer spacing $d = 0.335$ nm, and decay length $\delta = 0.184 a = 0.0453$ nm. The Dirac point then occurs at $K$ at an energy of $E_0=0.7833$ eV which is an overall shift we add to the Hamiltonian. The atomic positions are obtained through a rigid rotation of two honeycomb lattices through an angle $\theta(m,n)=\mathrm{Arg}[(m\omega^*+n\omega)/(n\omega^*+m\omega)]$ where $(m,n)$ parameterize a commensurate rotation and $\omega=e^{i\pi/6}$. The rotation center is a given by a point where two $A$ sublattice sites overlap (no displacement between layers), and this point can be chosen to be the corner of the unit cell.
	
	In Fig. S2, we plot the crystal momentum resolved spectral functions and densities of states calculated using the tight-binding model for all the samples' twist angles with the chemical potential illustrated by the dashed black line.
	
	\begin{center}
		\centering
		
		\includegraphics[width=0.265\linewidth]{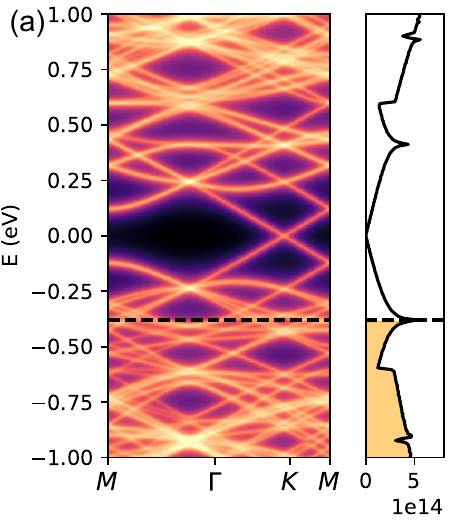}
		\includegraphics[width=0.23\linewidth]{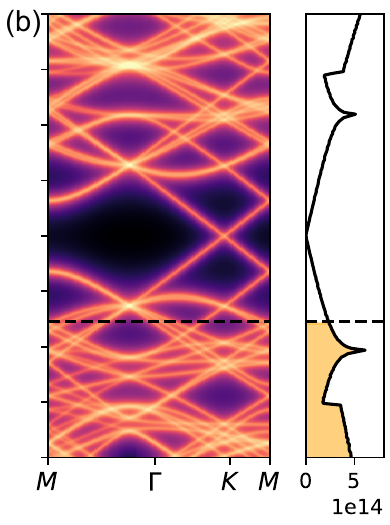}
		\includegraphics[width=0.23\linewidth]{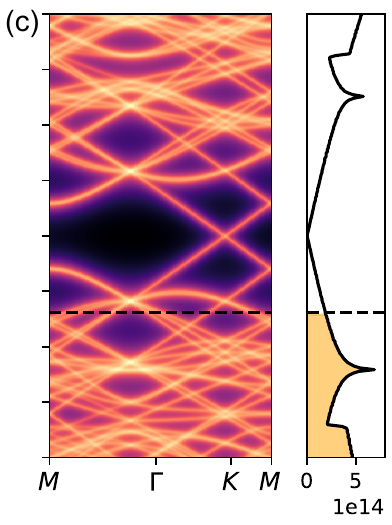}
		\includegraphics[width=0.23\linewidth]{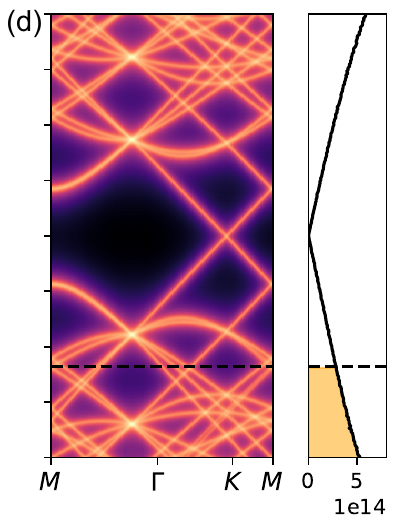}
		
		\small
		FIG. S2. Spectral functions and densities of states (in 1/eV cm$^2$) for samples with \textbf{(a)} $\theta(21,17)=6.96^\circ$, \textbf{(b)} $\theta(17,13)=8.80^\circ$, \textbf{(c)} $\theta(19,14)=10.00^\circ$, \textbf{(d)} $\theta(15,8)=19.93^\circ$ twist angles. The chemical potential for which the tight-binding model gives the same carrier density of states as the Hall effect measurements is indicated by the dashed black lines. We take $\eta=0.010$ eV for these calculations.
		\label{fig:dos}
	\end{center}
	
	While we see that while there are bands below the van Hove singularity other than the linearly dispersing bands originating from $K$, the DOS evolves smoothly with twist angle. This reflects the fact that these bands are either (1) back-folded linearly dispersing bands with some hybridization from Bragg scattering with the crystalline potential, or (2) back-folded from elsewhere in the zone where the coupling to the linearly dispersing bands is weak.
	
	\section{Extended Data}\label{section:data}
	
	\begin{table*}[]
		\centering
			\begin{tabular}{c|c|c}\hline\hline
				Angle ($^\circ$)& $n_h$ ($10^{13}/\mathrm{cm}^2$) & $m^*$ ($m_e$)\\ \hline
				7		& $4.89878\pm0.00148$	& $0.0570\pm 0.0033$\\
				9		& $3.96624\pm0.00078$	& $0.0728\pm 0.0035$\\
				10		& $3.08431\pm0.00078$	& $0.0558\pm 0.0033$\\
				20		& $7.86919\pm0.00642$	& $0.0986\pm 0.0018$
				\\\hline\hline
			\end{tabular}
			\hspace{0.15 in}
			\begin{tabular}{c|c|c|c}\hline\hline
				$\mu^\mathrm{dirac}$ (eV) & $v_F^\mathrm{dirac}$ ($10^6$ m/s) & $\mu^\mathrm{TB}$ (eV) & $v_F^\mathrm{TB}$ ($10^6$ m/s) \\ \hline
				$-0.514\pm0.029$	& $1.259\pm0.051$	& $-0.375\pm 0.029$ & $1.075\pm0.052$\\
				$-0.326\pm0.016$	& $0.887\pm0.030$	& $-0.376\pm 0.016$ & $0.953\pm0.030$ \\
				$-0.331\pm0.020$	& $1.020\pm0.043$	& $-0.344\pm 0.020$ & $1.041\pm0.043$\\
				$-0.478\pm0.009$	& $0.923\pm0.012$	& $-0.577\pm 0.009$ & $1.015\pm0.012$\\
				\hline\hline
		\end{tabular}
		
		\caption{\textbf{(left)} Extracted hole carrier density $n_h$ and cyclotron mass $m^*$ as dependent on twist angle. \textbf{(right)} Chemical potential and Fermi velocity derived from the extracted parameters at left using a continuum Dirac model and a full lattice model. Note the close agreement for the Dirac and tight-binding models for the $20^\circ$ sample which agrees with the analysis that these samples are well within in the Dirac regime.} \label{table:data}
	\end{table*}
	
	Here we provide the complete data necessary to support the conclusions drawn in the main text. The key parameters extracted from the experiments are tabulated in Table \ref{table:data}.
	
	The uncertainties for $n_h$ and $m^*$ are the square root of the covariance matrix elements (standard deviations) of the extracted fit parameters. For $\mu=(\pi\hbar^2/4)n_h/m^*$, the propagation of uncertainty gives
	\begin{align}
		\sigma_\mu &= \sqrt{(\partial_{n_h}\mu)^2\sigma_{n_h}^2 + (\partial_{m^*}\mu)^2\sigma_{m^*}^2}\\
		&=\mu \sqrt{\frac{\sigma_{n_h}^2}{n_h^2} + \frac{\sigma_{m^*}^2}{(m^*)^2}}
	\end{align}
	which we assume holds for both the Dirac model and the tight-binding model. For the propagation of uncertainty to $v_F=\sqrt{\mu/m^*}$ we have
	\begin{align}
		\sigma_{v_F} &= \sqrt{(\partial_\mu v_F)^2\sigma_\mu^2 + (\partial_{m^*}v_F)^2\sigma_{m^*}^2}\\
		&= \frac{v_F}{2}\sqrt{\frac{\sigma_\mu^2}{\mu^2} + \frac{ \sigma_{m^*}^2}{(m^*)^2}}
	\end{align}
	
\end{document}